# A Step-by-Step Guide to Using BioNetFit

William S. Hlavacek, Jennifer Longo, Lewis R. Baker, María del Carmen Ramos Álamo, Alexander Ionkov, Eshan D. Mitra, Ryan Suderman, Keesha E. Erickson, Raquel Dias, Joshua Colvin, Brandon R. Thomas and Richard G. Posner

**Running Head:** BioNetFit

**Corresponding Author:**

Richard G. Posner

Department of Biological Sciences

Northern Arizona University

Flagstaff, AZ 86011, USA

Email: richard.posner@nau.edu

# Summary

BioNetFit is a software tool designed for solving parameter identification problems that arise in the development of rule-based models. It solves these problems through curve fitting (i.e., nonlinear regression). BioNetFit is compatible with deterministic and stochastic simulators that accept BioNetGen language (BNGL)-formatted files as inputs, such those available within the BioNetGen framework. BioNetFit can be used on a laptop or standalone multicore workstation as well as on many Linux clusters, such as those that use the Slurm Workload Manager to schedule jobs. BioNetFit implements a metaheuristic population-based global optimization procedure, an evolutionary algorithm (EA), to minimize a user-defined objective function, such as a residual sum of squares (RSS) function. BioNetFit also implements a bootstrapping procedure for determining confidence intervals for parameter estimates. Here, we provide step-by-step instructions for using BioNetFit to estimate the values of parameters of a BNGL-encoded model and to define bootstrap confidence intervals. The process entails the use of several plain-text files, which are processed by BioNetFit and BioNetGen. In general, these files include 1) one or more EXP files, which each contains (experimental) data to be used in parameter identification/bootstrapping; 2) a BNGL file containing a model section, which defines a (rule-based) model, and an actions section, which defines simulation protocols that generate GDAT and/or SCAN files with model predictions corresponding to the data in the EXP file(s); and 3) a CONF file that configures the fitting/bootstrapping job and that defines algorithmic parameter settings.

# 1. Introduction

An inconvenient truth in systems biology is that models for cellular regulatory networks, with some exceptions (*see* **Note 1**), explicitly incorporate parameters that invariably and frustratingly have uncertain values. The parameters of a model are typically constants related to the physical properties of the system being represented by the model, such as initial abundances of the system's constituent parts or empirical rate constants that characterize the kinetics of processes of interest for particular conditions (e.g., physiological temperature). Parameter uncertainty is an important concern because model parameters influence model predictions, to varying degrees, as can be quantified via sensitivity analysis (Zi, 2011) *(1)*. Reliable predictions require reliable parameter estimates, for at least some subset of a model's parameters. (Predicted behaviors of concern may be insensitive to certain parameters.)

A factor that adds to the problem of parameter uncertainty is that some types of model parameters, such as an apparent equilibrium constant or a pseudo first-order rate constant, may have values that depend on factors outside the intended scope of a model, which limits the usefulness of many direct measurements (*see* **Note 2**). Even if direct measurements are feasible, it has been argued that parameter uncertainty is better addressed for systems biology models through curve fitting (Gutenkunst et al., 2007) *(2)*, in part because of measurement errors, which in combination can significantly degrade the quality of a model's predictions even if the errors are small (*see* **Note 3**).

Regardless of the value of direct parameter measurements, curve fitting is widely practiced, not only in systems biology but in many diverse fields, and it may be the only means available to estimate at least some of a model's parameters. Fitting involves adjusting a model's parameter values, perhaps within specified box constraints (e.g., a minimum and a maximum that define a feasible range), so that model predictions match (as best as possible according to a specified metric for goodness of fit) available data characterizing relevant system behavior. The data used in curve fitting commonly take the form of time courses or (steady-state) dose-response curves. The focus is typically on quantitative data (i.e., numerical data), but *qualitative* data, such as categorizations, may also be used to constrain parameter estimates, either alone or in combination with quantitative data (Mitra et al., in press) *(3)*.

There are many software tools available that enable curve fitting for models having traditional forms, such as that of ordinary differential equations (ODEs). An example is Data2Dynamics (Raue et al., 2015) *(4)*, which is designed for the analysis of ODE models. In contrast, there are relatively few curve-fitting tools designed for compatibility with simulators that enable rule-based modeling (*see* **Note 4**), such as those available within the BioNetGen framework (Blinov et al., 2004; Harris et al., 2016) *(5, 6)*, which encompasses BioNetGen, NFsim (Sneddon et al., 2011) *(7)*, and RuleBender (Xu et al., 2011; Smith et al., 2012) *(8, 9)*, or the Kappa platform (Boutillier et al., 2018) *(10)*. Examples of tools in this class include RKappa (Sorokin et al., 2015) *(11)* and BioNetFit (Thomas et al., 2016) *(12)*.

BioNetFit is a freely available open-source software package for estimating parameter values of rule-based models defined using the BioNetGen language (BNGL) (Faeder et al., 2009; Hogg et al., 2014) *(13, 14)*. BNGL and related languages (e.g., Kappa) are designed for the definition of models that track biomolecular site dynamics (Chylek et al., 2014a) *(15)*, i.e., models that capture the dynamics of biomolecular site state transitions (*see* **Note 5**). BioNetFit can find a global fit to multiple datasets via an evolutionary algorithm, which is a type of population-based metaheuristic optimization algorithm. For a survey of this class of algorithms, see Boussaïd et al. (2013) *(16)*. The algorithm implemented in BioNetFit 1 is outlined in the Supplementary Methods section of the report of Thomas et al. (2016) *(12)*. To enable the quantification of parameter uncertainty, BioNetFit also implements a bootstrapping procedure (Efron and Tibshirani, 1994; Press et al., 2007) *(17, 18)*.

BioNetFit is compatible with both the deterministic and stochastic simulation engines available within the BioNetGen framework (Blinov et al., 2004; Harris et al., 2016) *(5, 6)*, including NFsim (Sneddon et al., 2011) *(7)*, which implements a network-free stochastic simulation algorithm (this methodology is briefly discussed in the paragraph below), and the run_network interface to CVODE (Hindmarsh et al., 2005) *(19)*, an ODE solver. Noisy simulation data produced by a stochastic simulation algorithm (SSA) can be problematic when fitting; simply put, this noise can prevent an optimization algorithm from converging. BioNetFit is designed to cope with this issue through smoothing (*see* **Note 6**).

There are at least two reasons that a modeler may wish or need to use an SSA. The most common reason is an interest in fluctuations arising from small population sizes (Munsky et al., 2012) *(20)*. As recently discussed by Suderman et al. (2018) *(21)*, an entirely different reason, which arises in rule-based modeling, is the need to cope with combinatorial complexity (Hlavacek et al., 2003; 2006) *(22, 23)* (*see* **Note 7**). In some cases, it may be impracticable or impossible to process the rules of a model to obtain an equivalent formulation in terms of a list of individual reactions or a coupled system of ODEs. In these cases, methods called network-free simulation algorithms, which are related to Gillespie's SSA (Gillespie, 2007) *(24)*, may be used to perform simulations (Suderman et al., 2018) *(21)*. In network-free simulation, the rules of a model serve as reaction event generators in a kinetic Monte Carlo (KMC) procedure (Danos et al., 2007; Yang et al., 2008) *(25, 26)*. For this reason, network-free methods have also been called direct methods—the rules of a model are used directly to perform a simulation. In contrast, in indirect methods, rules are processed to first obtain an equivalent model in a traditional form (e.g., a system of coupled ODEs), and then that derivative form of the original model is used to perform a simulation (e.g., through numerical integration of the rule-derived ODEs) (Faeder et al., 2005; Blinov et al., 2006) *(27, 28)*.

Although network-free methods enable simulations when the rules of a model imply a reaction network that would be impracticable to derive, these methods sometimes have the drawback of being computationally inefficient, because system state is advanced one reaction event at a time. For this reason, and also because of the computational challenges of optimization, especially in high-dimensional parameter spaces, another important feature of BioNetFit is its ability to leverage parallel computing resources, such as multicore workstations and Linux clusters.

In this chapter, we provide a step-by-step guide for setting up and running a fitting job with BioNetFit. We also elaborate on how to run a bootstrapping job, which entails essentially the same steps with only minor changes. We intentionally avoid a discussion of the full range of BioNetFit's features—for these details, we refer the reader to the BioNetFit user manual, which is included in the BioNetFit distribution. Instead, we focus on the steps required to execute specific fitting and bootstrapping jobs. It is hoped that the recipes given here provide helpful illustrations that can guide other use cases. Two benefits of BioNetFit for modelers who are using BNGL or Kappa to define models (*see* **Note 8**) are that 1) there is no need to reinvent scripts for running fitting and bootstrapping jobs and 2) parameter identification and definition of confidence intervals become more reproducible.

## 2. Materials

The BioNetFit source code is freely available online (https://github.com/RuleWorld/BioNetFit) *(29)*; it is distributed under the terms of the GNU General Public License v3.0 *(30)*. The application file is named "BioNetFit.pl." As the name implies, BioNetFit is a Perl script/program and, consequently, it is compatible with laptops and workstations on which Perl is installed, including macOS, Windows/Cygwin, and Linux platforms. In addition, BioNetFit can be readily deployed on many Linux-based clusters. It is compatible with two different cluster management and job scheduling systems: the Slurm Workload Manager and TORQUE (Thomas et al., 2016) *(12)*. BioNetFit has been most extensively tested on two clusters at Northern Arizona University and Los Alamos National Laboratory, called Monsoon and Darwin, which are Slurm clusters. Use of BioNetFit requires an installation of the BioNetGen framework. The BioNetGen source code is freely available online (https://github.com/RuleWorld/bionetgen) *(31)*; it is distributed under the terms of the MIT License *(32)*. Conveniently, precompiled binaries are also available for download for a variety of platforms at the Faeder Lab web site (https://www.csb.pitt.edu/Faculty/Faeder/) *(33)*. Because BioNetFit is a command-line tool, its use requires a terminal emulator (e.g., the Terminal.app utility available on the macOS platform).

    Detailed installation advice is provided in the BioNetFit user manual, which is a PDF file named "BioNetFit_User_Manual.pdf." This file is available online (http://bionetfit.nau.edu/files/BioNetFit_User_Manual.pdf) *(34)* and is included in the software distribution package. The distribution also includes multiple sets of EXP, BNGL, and CONF files (in the "examples" directory/folder) needed to execute example fitting and bootstrapping jobs. If technical support is required, we recommend sending an email request for help to bionetgen.help@gmail.com, which will reach the community of software developers working on BioNetGen and BioNetFit.

Below, we will walk through the steps required to execute specific fitting and bootstrapping jobs. During this discussion, it will be helpful to understand how BioNetFit and BioNetGen interact with input and output files; the relationships between software components and input/output files are illustrated in **Fig. 1**. The files needed to execute the fitting/bootstrapping jobs that we will discuss are available online *(35, 36)*. These files and the usage instructions given below are tailored for BioNetFit version 1. Although we are continuing to develop BioNetFit—the code is being refactored to add a variety of new features—the usage instructions below should remain largely relevant for the foreseeable future.

## 3. Methods

### 3.1. Setting up a fitting job

A fitting job requires (and is defined by) several problem-specific plain-text files: 1) a BNGL file (*see* **Note 9**), which should define a) the model of interest, b) the model parameters that will be free to vary during fitting, c) simulation outputs corresponding to experimental readouts of interest, and d) appropriate protocols for generating the relevant simulation data; 2) one or more EXP files with the experimental data to be used in fitting (viz., time courses and/or dose-response curves); and 3) a CONF file, which controls the behavior of BioNetFit. Below, we will focus on a specific fitting problem, which we will refer to as the egg fitting problem. We will also briefly discuss a more computationally challenging version of this problem, which we will refer to as the elephant fitting problem, and a bootstrapping problem, which can be viewed as an extension of the egg fitting problem.

For purposes of illustration, and to keep the model under consideration simple, we will consider a toy fitting problem involving the use of a truncated Fourier series expansion for each of two coordinates to find a mathematical representation of a closed contour, as in elliptic Fourier analysis (EFA) (Kuhl and Giardina, 1982) *(37)*. More specifically, the problem is to find Fourier descriptors for the two-dimensional contour of an egg (**Fig. 2A**) via BioNetFit-enabled optimization (*see* **Note 10**). In other words, we will attempt to find coefficients $a_0, c_0, a_n, b_n, c_n,$ and $d_n$ $(n = 1, ..., N)$ in the following equations that allow for an accurate representation of a contour of interest:

$$X_N(t) = a_0 + \sum_{n=1}^{N} a_n \cos\left(\frac{2n\pi t}{T}\right) + b_n \sin\left(\frac{2n\pi t}{T}\right) \quad (1)$$

$$Y_N(t) = c_0 + \sum_{n=1}^{N} c_n \cos\left(\frac{2n\pi t}{T}\right) + d_n \sin\left(\frac{2n\pi t}{T}\right) \quad (2)$$

Here, each $(X_N(t), Y_N(t))$ pair can be interpreted as referencing a position (in a Cartesian coordinate system) on the contour of interest at a time $t$ (measured in units of, say, s) as the contour is traversed at constant velocity with period $T$. We will take $N = 2$ and $T = 180$ s.

The BNGL, EXP and CONF files that define the toy fitting problem (namely, egg.bngl, egg.exp, and egg_fit.conf) are available online *(35)*. Creating these files comprises the work of setting up our example BioNetFit fitting job. Each file is discussed below.

### 3.1.1. The EXP file

The EXP file egg.exp, which is illustrated in **Fig. 2B**, contains coordinates in a tabular format for a sampling of 180 points from the contour of **Fig. 2A**. To generate this file, we recorded the $(X, Y)$ coordinates, with imprecision, at every second during a constant-velocity traversal of the contour beginning at $t = 0$. Thus, the points in egg.exp are roughly evenly spaced, such that the arc length between adjacent points is the similar for all such points. The mean arc length is 1.6 (in the same units as $X$ and $Y$) and the standard deviation is 0.47. Note that the points at $t = 0$ and 180 s are identical, in harmony with the period of traversal that we have chosen, $T = 180$ s.

The format of an EXP file (**Fig. 2B**) is the same as that of a GDAT or SCAN file (Faeder et al., 2009) *(13)* (*see* **Note 11**). Thus, the first line of an EXP file must begin immediately with a number sign (#). The next header must be the name of the independent variable: "time" for a time course or the name of a parameter for a dose-response curve (corresponding to a parameter scan defined by an action in a BNGL file) (*see* **Note 12**). Additional headers should correspond to the names of observables and/or global functions defined in the BNGL file corresponding to the EXP file; order is unimportant. Headers (and columns of data) should be separated by white space. After the first row, each line in an EXP file should contain numerical data, with one entry for each header.

For our toy fitting problem, only a single EXP file is required, but multiple EXP files are allowed. For other problems, experimental data may be placed in a single EXP file or distributed among multiple EXP files. Use of multiple EXP files is required if a user is seeking a global fit to mixed data types (i.e., data including both time courses and dose-response curves) or multiple dose-response curves corresponding to scans of different parameters. In general, EXP files should be mappable to actions defined in a corresponding BNGL file. In other words, the simulation outputs generated by a given action should relate to the experimental data in a corresponding EXP file. Each EXP file to be used in a fitting job must be identified by path and filename in the CONF file associated with the fitting job, as we will discuss below in further detail.

### 3.1.2. The BNGL file

We assume that the reader is familiar with BNGL (Faeder et al., 2009; Hogg et al., 2014) *(13, 14)* (*see* **Note 13**). If not, tutorial introductions are available elsewhere (Sekar and Faeder, 2012; Chylek et al., 2015) *(38, 39)*.

The BNGL file for the egg fitting problem (egg.bngl) is shown in **Fig. 3A**. This file is setup to calculate $(X_2(t), Y_2(t))$ for $t = 0, 1, \ldots, 180$ s on the basis of **Eqs. 1** and **2** (with $N = 2$ and $T = 180$ s). Lines 3–12 in the BNGL file (**Fig. 3A**) identify the coefficients $a_0, a_1, a_2, b_1, b_2, c_0, c_1, c_2, d_1$, and $d_2$ in **Eqs. 1** and **2** as the parameters that will be allowed to vary in fitting. For given values of these coefficients and a time $t$, Lines 27–32 (**Fig. 3A**) are dedicated to calculating the coordinates of a point on the contour of **Fig. 2A**. The (synthesis) rule on Line 35, in combination with the commands in the actions section of the BNGL file (Lines 38–43), serves to increment time.

The most important feature of the BNGL file to notice is the mapping of parameter names to labels that each end with "`__FREE`" (see Lines 3–12, **Fig. 3A**). These labels are also found in the corresponding CONF file (see Lines 37–46, Fig. 3B). Ordinarily, parameters introduced in the parameters block of a BNGL file would be assigned numerical values or mapped to mathematical expressions, as on Lines 13–15 (**Fig. 3A**). During a fitting run, BioNetFit will repeatedly replace the labels ending with "`__FREE`" in the BNGL file with trial parameter values.

A second feature of the BNGL file to notice is the definition of simulation outputs, the global functions `X()` and `Y()` (on Lines 27–32, **Fig. 3A**). Importantly, these functions have names corresponding to headers in the EXP file (**Fig. 2B**). Note that the `print_functions=>1` setting in the `simulate` command (see Line 42, **Fig. 3A**) instructs BioNetGen to report global function values in the output GDAT file; this overrides the default behavior, which is to omit function values. In the GDAT file generated as BioNetGen processes egg.bngl, the relevant simulation outputs for the egg fitting problem are the time-varying values of the global functions `X()` and `Y()`. In general, simulation outputs may be defined using any combination of global functions and observables. In this particular GDAT file, an observable named `t` is reported. This observable is not an output that will be directly compared to data; it is introduced only to allow time to enter into function evaluations. Extra columns in a GDAT file (i.e., columns not corresponding to columns in an EXP file) are ignored by BioNetFit.

A third feature of the BNGL file to notice is that the times selected for reporting simulation results match the times listed in the corresponding EXP file (**Fig. 2A**). Report times are defined by the settings `t_start=>0`, `t_end=>180`, and `n_steps=>180` in the `simulate` command (see Line 41, **Fig. 3A**). These settings direct BioNetGen to report simulation results for $t = 0, 1, \ldots, 180$. Any extra report times are ignored by BioNetFit.

Finally, we note that the `suffix` setting in the `simulate` command (Line 40, **Fig. 3A**) is important. It should match the base name of the EXP file containing the experimental data that will be compared against the simulation data to be generated by the `simulate` command, which here is "egg."

The features noted above illustrate how a BNGL file should be setup to produce simulation outputs that are comparable to the experimental data being used in fitting. Outputs need to have names that match headers in EXP files and, in the case of time-course data, report times need to include the times listed in EXP files.

### 3.1.3. The CONF file

The behavior of BioNetFit is controlled by a user-supplied CONF file, in which BioNetFit parameters are mapped to desired settings (*see* **Note 14**). Optional parameters that are not explicitly set in a user-supplied CONF file take on their default settings. An error report is generated if a CONF file does not include a required setting (e.g., a path to a BNGL file or a path to at least one EXP file). See the BioNetFit user manual *(34)* for a full listing of all parameters and their default settings (if any).

The CONF file for the egg fitting problem (egg_fit.conf) is shown in **Fig. 3B**. At the top of this file, settings are provided for the BioNetFit parameters `output_dir`, `job_name`, `bng_command`, `model`, and `exp_file`. These settings simply define paths and directory/file names, as indicated in the comment lines. (Note that each comment line begins with a number sign.) The string assigned to `job_name` defines the name of a directory where results will be sent; this directory is taken to be a subdirectory of the directory identified by the `output_dir` setting. Note that paths may be absolute or relative to the current working directory.

On Line 21 in the CONF file, the parameter `parallel_count` is assigned an integer value of 1, which indicates that a single processor should be dedicated to execution of the fitting job. BioNetFit can take advantage of multiple processors if more than one is available. On Line 24 in the CONF file, the parameter `objfunc` is assigned a value of 1, which indicates that a sum-of-squares function should be used to evaluate quality of fit *(34)*. On Lines 27, 31, and 34, the parameters `max_generations`, `permutations`, and `mutate` are assigned values; these are parameters of the optimization algorithm. The value of `max_generations` is the number of desired iterations of the algorithm. The value of `permutations` is the number of trial sets of parameter values to be considered at each iteration. The settings for `mutate` determine how parameter values are to be varied from iteration to iteration. A detailed discussion of the algorithm is beyond the intended scope of this tutorial. If these details are of interest, the reader is referred to the report of Thomas et al. (2016) *(12)* and the BioNetFit user manual *(34)*.

Lines 37–46 of the CONF file identify the parameters that will be free to vary during fitting and initialize the search of parameter space. Here, the coefficients in **Eqs. 1** and **2** for *N* =

2 (which are named `a0`, `a1`, `a2`, etc. in the BNGL file) are identified as the free parameters and a range is given for each. Because it is known that $(a_0, c_0)$ corresponds to the coordinates of the center of mass of a contour being represented by **Eqs. 1** and **2** (Kuhl and Giardina, 1982) *(37)*, we have assigned ranges for $a_0$ and $c_0$ that define a box that is tightly centered on the center of mass of the contour of **Fig. 2A**, which can be roughly identified by inspection. For each of the other coefficients, we assign a broader range, from –50 and 50. The initial trial values for each of these parameters, which will be chosen randomly, will lie within this range. Various options are available for initializing the search of parameter space; the reader is referred to the BioNetFit user manual *(34)* for further discussion.

### 3.1.4. Summary of the steps involved in setting up a fitting job

In summary, and in general, the steps involved in setting up a fitting job are as follows:

1. Place the available experimental data of interest in one or more EXP files. EXP files are plain-text files (with the filename extension .exp) that have the same tabular format of a GDAT or SCAN file (Faeder et al., 2009) *(13)*. The format is exemplified in **Fig. 2B**. The first row must begin with the number sign (#) in the first column and headers should conform with the syntax of BNGL parameter names (Faeder et al., 2009) *(13)*. These names must start with a letter; they may not contain white spaces or other non-alphanumeric characters with the exception of the underscore (_) character.

2. Modify the BNGL file that defines the model of interest (as needed) for consistency with the EXP file(s). Appropriate simulation outputs with names corresponding to headers in the EXP file(s) should be defined in the observables block and/or functions block of the model section. Furthermore, appropriate protocols for generating relevant simulation data should be defined in the actions section. Recall that a BNGL file is plain-text file that ends with the filename extension .bngl.

3. Modify the BNGL file for consistency with the CONF file that will be used to control the behavior of BioNetFit. This step involves removing any parameter settings for the parameters that will be allowed to vary in the fitting procedure (i.e., the free parameters). The free parameters should then be identified by mapping each of their names to a label ending with "__FREE" as illustrated in **Fig. 3A** (at the top of the `parameters` block). The labels should match those listed in the CONF file (cf. Lines 3–12 in **Fig. 3A** and Lines 37–46 in **Fig. 3B**). In other words, for a free parameter with the name *parameterName*, the label should take the form *parameterName*__FREE. These strings, *parameterName* and *parameterName*__FREE, are associated with each other by including a line in the BNGL file that begins with the parameter name, which is then followed by white space and finally the corresponding label.

4. Create a CONF file, which is a plain-text file that ends with the filename extension .conf. The content of the CONF file will depend on problem-specific requirements/preferences. Recall that a complete listing of a CONF file (for the egg fitting problem) is shown in **Fig. 3B** and that other examples are included in the BioNetFit distribution (Thomas et al., 2016) *(12, 29)*. In addition, the CONF files for the egg and elephant fitting problems are available online *(35, 36)*. The important elements of a CONF file include a) path settings, b) hardware/environment settings (e.g., the number of processors available to be used in parallel), c) algorithmic parameter settings, and d) a listing of labels used in the corresponding BNGL file to mark the parameters that will be allowed to vary during the fitting procedure. These labels are used in the CONF file to declare how the search of parameter space will be initialized.

5. Place the EXP, BNGL and CONF files together in a directory/folder that has a name related to the job name defined in the CONF file.

Recall that additional guidance is available in the BioNetFit user manual *(34)* and that the developers of BioNetGen and BioNetFit can be reached via email if technical support is needed (*see* **Materials**).

## 3.2. Running a fitting job

### 3.2.1. Consistency of path settings with file locations

Before starting the egg fitting job, you should ensure that the paths set in your CONF file (e.g., egg_fit.conf) using `bng_command`, `model`, and `exp_file` (see **Fig. 3B**) are consistent with the actual file system locations of, respectively, BNG2.pl (the Perl component of BioNetGen), your BNGL file (egg.bngl), and your EXP file (egg.exp). (Multiple EXP files may be identified in a CONF file by including multiple `exp_file` settings, one for each file.) You should also ensure that the settings for `output_dir` and `job_name` provide the desired names/locations of the output directory and its job-specific subdirectory, which will be created by BioNetFit if necessary and updated with content changes as your fitting job progresses. The content of existing output directories may be overwritten, if desired.

### 3.2.2. How to start a fitting job at the command prompt

BioNetFit is a command-line tool. Thus, you will need a terminal emulator, such as the Terminal.app utility available within the macOS platform, to use BioNetFit. At the command prompt of your terminal emulator, use the following command to start a fitting job:

```
perl /path/to/BioNetFit.pl /path/to/egg_fit.conf
```

Of course, you should replace the instances of `/path/to` with the proper paths to BioNetFit.pl and the relevant CONF file (egg_fit.conf). Note that relative paths defined in a CONF file are relative to the working directory. Thus, if you first navigate to the BioNetFit_v1.01 directory, you can start a job for the egg fitting problem with the command given on Line 3 (a comment line) in egg_fit.conf (**Fig. 3B**).

In addition to the CONF file path command-line argument (`/path/to/egg_fit.conf`), other command-line arguments available are `results`, `resume`, and an integer (which overrides the setting of `max_generations` in the CONF file being used). The `results` argument will generate a progress report. The `resume` argument will restart a fitting job. This argument is useful if a job is interrupted. It is also useful, when combined with an integer argument, for continuing a fitting job for a greater number of iterations than what was originally requested in a CONF file. An example command that includes `resume` and the integer argument is shown below (at Step 5 in Section 3.2.4).

### 3.2.3. Where to find the results of fitting

For the egg fitting problem, final results will be found in the "egg_out/egg_fit/Results" directory, which is a subdirectory of a subdirectory of the output directory identified in the CONF file, on Lines 6 and 9 of egg_fit.conf (**Fig. 3B**). Note that the "egg_out" directory will be found within the same directory where you started the fitting run (e.g., within BioNetFit_v1.01 if this was the working directory when you started the fitting run). You will want to inspect two files in the "egg_out/egg_fit/Results" directory: a BNGL file and a GDAT file.

The output BNGL file will have a name of the form egg_perm[*integer*].bngl, where the integer is the index for one of the trial parameter sets considered in fitting. The output file is a slightly modified version of the user-supplied BNGL file (egg.bngl). The file has been changed to include a definition of the best-fit values for the free parameters. In other words, a setting for each parameter with the suffix "__FREE" (from the best-fitting set of parameter values) has been added at the top of the parameters block, immediately below the `begin parameters` line:

```
a0__FREE   9.98588200e+01
a1__FREE  -1.5999241385921
a2__FREE   1.75445028e+00
b1__FREE  -3.04534132e+01
b2__FREE   1.17265139e+00
c0__FREE   1.39457620e+02
c1__FREE  -3.91167731e+01
c2__FREE  -5.43746173e-01
d1__FREE   1.87269903e+00
d2__FREE   1.75650430e+00
```

With the addition of these lines, the BNGL file becomes executable, because now, all parameters have well-defined numerical values. It should be noted that the original, user-supplied BNGL file (egg.bngl) cannot be processed directly by BioNetGen, because `a0, a1, a2`, etc. are not assigned numerical values in the file. In contrast, in the output BNGL file, `a0` is given the value assigned to `a0__FREE`, and so on. Best-fit values for $a_0, c_0, a_n, b_n, c_n$, and $d_n$ for $n = 1$ and 2 are summarized in Table 1; for comparison, this table also gives the values determined using the methodology of Kuhl and Giardina (1982) *(37)*. These latter values yield a reasonable reconstruction of the original egg shape, whereas the values found via optimization yield a less-perfect reconstruction. The quality of the fit found using BioNetFit is illustrated in **Fig. 4**.

The other output file of special interest is named "egg_bestfit.gdat." This file simply reports the results of a simulation run performed on the basis of the best-fit parameter values. The data recorded in this file were used to construct **Fig. 4**.

### 3.2.4. Summary of the steps involved in running a fitting job

In summary, and in general, the steps involved in running a fitting job (on a laptop or workstation) are as follows:

1. Ensure that files (e.g., the EXP, BNGL and CONF files that define your fitting job) are in the proper locations on your file system. You will need a laptop or workstation with Perl and BioNetGen installed.

2. Using a terminal emulator, execute the BioNetFit.pl program at the command line prompt with the path to your CONF file as a command-line argument. An example of a command that will start a fitting job is

   ```
   ./BioNetFit.pl egg/egg_fit.conf
   ```

   This command is appropriate if the working directory is the BioNetFit directory (which contains the Perl program BioNetFit.pl) and if a) the input directory for the egg fitting job is named "egg," b) this directory is a subdirectory of the BioNetFit directory, and c) this directory contains the files egg.exp, egg.bngl, and egg_fit.conf. Note that, before execution begins, you may need to respond to a prompt asking if you would like to overwrite existing results.

3. After your fitting job terminates successfully, the final results are stored in several files that are written to a subdirectory of a subdirectory of the output directory. The output directory is that named in the CONF file (see Line 6 in egg_fit.conf, **Fig. 3B**). If this directory does not already exist, it will be created. The subdirectory of the output directory that is of interest will have the job name defined in the CONF file (e.g., "egg_fit" because of `job_name=egg_fit` at Line 9 in egg_fit.conf, **Fig. 3B**). The subdirectory of this subdirectory, where final results are stored, is named "Results." The output files in Results include a) a modified copy of the user-supplied CONF file with, for example, additions made for default settings, b) a BNGL file with free parameters set to their best-fit values, c) one or more GDAT and SCAN files with data from simulation runs with best-fit parameter values, and d) a file named "sorted_params.txt." This last file provides a sorted listing of objective function values for the various sets of trial parameter values considered during fitting. Intermediate results, which are important for restarts, are stored in a series of subsubdirectories having integers as names, one for each iteration of the optimization algorithm. The integer names are the indices of the iterations.

4. Evaluate the quality of fit by comparing the data in your EXP file(s) against the data in the GDAT/SCAN files produced through simulation(s) with the best-fit parameter values.

5. If the quality of fit is unacceptable, you might be able to improve the fit by executing additional iterations of the optimization algorithm. A fitting run can be restarted by using a command such as

    `./BioNetFit.pl resume 100 egg/egg_fit.conf`

    where the command-line argument `resume` simply indicates that execution is to resume from the last stopping point and the command-line argument `100` defines a new setting for the algorithmic parameter `max_generations` originally defined in the CONF file (see Line 27 in **Fig. 3B**). The new setting should exceed the original setting.

### 3.3. Limitations of fitting

BioNetFit is not a panacea. Fitting problems may be difficult, if not impracticable or even impossible, to solve. This point is illustrated with the elephant fitting problem, which is defined by the files elephant.exp, elephant.bngl, and elephant_fit.conf, which are available online *(36)*. The elephant fitting problem is similar to the egg fitting problem, except that the input contour data defines a more complicated shape, the life-like silhouette of an elephant (**Fig. 5A**). A faithful EFA representation of this shape requires in excess of 30 harmonics (i.e., $N > 30$); however, a reasonable approximate representation is possible for $N = 20$ (**Fig. 5B**). Although the elephant fitting problem is configured to start near the point in parameter space that provides a reasonable approximation (determined by EFA), BioNetFit is unable to converge (in a reasonable amount of time) to a fit having the same quality as the EFA representation of **Fig. 5B**, as can be seen in **Fig. 5C**. The problem is that the parameter space being searched has $4N + 2 = 82$ dimensions. Thus, BioNetFit fails to find a good solution because of the curse of dimensionality. Unfortunately, there is no generally applicable prescription for addressing the limitations of fitting. Experimental design for the purpose of generating informative data that can reduce parameter uncertainty (Apgar et al., 2010; Tönsing et al., 2014) *(40, 41)* would likely be helpful in many circumstances.

### 3.4. Bootstrapping

#### 3.4.1. The boostrapping algorithm implemented in BioNetFit

We now turn our attention to the bootstrapping feature of BioNetFit, which provides a means to determine confidence intervals for parameter estimates. Bootstrapping is a resampling procedure that entails the random generation of pseudo datasets from an available empirical dataset (Efron and Tibshirani, 1993; Press et al., 2007) *(17, 18)*. Various bootstrapping and related resampling procedures (e.g., jackknifing) have been described in the literature; BioNetFit implements the algorithm described by Press et al. (2007) *(18)*. In this procedure, the set of *N* experimental data points recorded in the EXP file(s) of a fitting job, which we denote as $(D_{(0)}^S)$, is randomly sampled with replacement to create a collection of *M* pseudo datasets $(D_{(1)}^S, \ldots, D_{(M)}^S)$, each of which also contains *N* data points. These pseudo datasets are often referred to as bootstrap samples. Sampling with replacement means that the pseudo datasets differ from the original dataset: in each pseudo dataset, some of the original data points may be missing and some may be represented multiple times. The effect of resampling is to give the original data points differing weights. After pseudo datasets have been generated, a fitting run is executed for each of the pseudo datasets (*see* **Note 15**). These runs are analogous to an ordinary fitting run; the only difference is the use of EXP file(s) derived from the original EXP file(s) through resampling. The end product of each fitting run is a set of parameter estimates. The resulting *M* sets of parameter estimates provide a statistical characterization of parameter uncertainty given the model of interest and the available data in the original EXP file(s) (Efron and Tibshirani, 1993) *(17)*. A minimum of 1,000 bootstrap samples has been recommended (Efron and Tibshirani, 1993) *(17)*, with more samples yielding a more precise statistical characterization of uncertainty.

### 3.4.2. Steps involved in running a bootstrapping job

A BioNetFit bootstrapping job, which is similar to a fitting job, involves the steps described below. It is assumed that EXP, BNGL and CONF files defining a fitting job are available.

1. There is no need to modify the EXP or BNGL files of the fitting job. All adjustments needed to switch from fitting to bootstrapping can be made by editing the CONF file of the fitting job. Compare the egg_fit.conf and egg_boot.conf files for the egg fitting and the egg bootstrapping problems, which are available online *(35)*. To change a CONF file that defines a fitting job to one that defines a bootstrapping job, one may simply add a few lines, as described below.

- Add a line that requests a specified number of bootstrap samples (e.g., `bootstrap=1000`).

- Add a line that sets an upper threshold for the objective function value (e.g., `bootstrap_chi=2.0`) (*see* **Note 16**). This threshold defines a maximum acceptable objective function value, which is used to automatically prune away results from fitting runs that do not produce an acceptable quality of fit (as determined by the specified threshold setting). Results from a failed fitting job are rejected and the fitting job is restarted (from a new starting point, if a randomized initialization of the search of parameter space is being used, as recommended). This feature is useful for a rugged fitness landscape, wherein the fitting procedure may become trapped in a local minimum that is far away from the global minimum.

- Add a line to define the number of attempts that will be made to satisfy the constraint defined by the `boostrap_chi` setting (e.g., `bootstrap_retries=5`). If no attempt succeeds, the bootstrapping procedure continues without further consideration of the problematic resampled dataset. If a setting for `bootstrap_retries` is not given, this parameter takes on its default setting.

2. Start the bootstrapping job just as you would a fitting job; the main difference is the command-line argument that gives the name of the CONF file to be processed by BioNetFit. For example, the egg bootstrapping job is started with a command such as `./BioNetFit.pl egg/egg_boot.conf`, whereas the egg fitting job is started with a command such as `./BioNetFit.pl egg/egg_fit.conf`.

3. After the bootstrapping procedure finishes, results (a collection of parameter estimates, with one for each bootstrap sample considered in the procedure) can be found in a plain-text file named "params.txt." This file will be located where specified in the CONF file for the boostrapping job. For the egg bootstrapping job, intermediate results can be found in the directory "egg_out/egg_boot" and the "params.txt" file can be found in the directory "egg_out/egg_boot_boostrap." The name of the latter directory is derived by adding "_bootstrap" to the end of the string provided with the `job_name` setting.

Importantly, in most circumstances, one should run the fitting jobs of a bootstrapping problem in parallel to the extent possible. Bootstrapping will usually involve numerous fitting runs, and thus, the ability of BioNetFit to leverage parallel computing resources becomes especially important. The egg bootstrapping job is configured to use multiple CPUs on a multicore workstation, which dramatically reduces the amount of time required to complete the bootstrapping job (compared to serial execution of the fitting runs on a single processor).

### 3.4.3. Calculating parameter confidence intervals

A confidence interval, defined by a lower bound $CI_{lower}$ and an upper bound $CI_{upper}$, can be calculated from the parameter estimates reported in the output file "params.txt" using the bootstrap percentile interval method:

$$(CI_{lower}, CI_{uppper}) = (\theta_{lower}, \theta_{upper})$$

where $\theta_{upper}$ is the $M(1 - \alpha/2)$ ranked bootstrap estimate, $\theta_{lower}$ is the $M(\alpha/2) + 1$ ranked bootstrap estimate, $\alpha$ is the desired confidence level (e.g., $\alpha = 0.1$ for a 90% confidence interval), and $M$ is the total number of bootstrap samples.

For example, if we have $M = 1,000$ bootstrap samples and want to identify the 90% confidence interval for a parameter, we simply sort all of the estimates for each parameter (independently) in order of increasing value, then select the 51st ranked estimate as the lower bound of the confidence interval and the 950th ranked estimate as the upper bound of the confidence interval.

### 3.5. Concluding remarks

The first release of BioNetFit (Thomas et al., 2016) *(12)* was mainly intended to serve as a prototype, a first step toward a powerful and robust tool tailored for rule-based modeling applications. However, it is capable of solving real-world fitting problems. The fitting problems solved originally by Kozer et al. (2013) *(42)* and Chylek et al. (2014c) *(43)* using purpose-built codes were redone as demonstrations (Thomas et al., 2016) *(12)*, and BioNetFit has since been applied to solve various original fitting problems (Mahajan et al., 2017; Harmon et al., 2017; Erickson et al., submitted) *(44–46)*.

BioNetFit is being actively developed, as an open-source project. Contributions (and reuses) of code are welcomed, and the developer community is available to help prospective users apply BioNetFit in their research (*see* **Materials** for the developer email address). In addition, usage of BioNetFit is now taught in the Annual q-bio Summer School (Resnekov et al., 2014) *(47, 48)*. We encourage users to submit EXP, BNGL and CONF files that define fitting jobs to the RuleHub repository *(49)*, where the files for the problems discussed here can be found *(35, 36)*. Contributions of files in the SBML multi format (Zhang and Meier-Schellersheim, 2018) *(50)* would also be welcomed, as support for this format within the BioNetGen framework is forthcoming. The RuleHub repository includes sections for both published and contributed (i.e., unpublished) files.

Future releases of BioNetFit will provide 1) a toolbox of metaheuristic optimization methods, including algorithms designed to leverage parallel computing resources (Egea et al., 2009; Penas et al., 2015) *(51, 52)*; 2) support for libRoadRunner (Somogyi et al., 2015) *(53)*, which will make BioNetFit compatible with models that can be encoded in the SBML format (Hucka et al., 2018) *(54)*; and 3) implementation of methods discussed by Mitra et al. (in press) *(3)* for using both quantitative and qualitative data to drive parameter identification. Despite these and other changes, the work flows presented here should remain relevant as guides for applying BioNetFit in the development of rule-based systems biology models for the foreseeable future.

## 4. Notes

1. An example of a model without explicitly defined parameters is a logical or Boolean model, such as the model of Miskov-Zivanov et al. (2013) *(55)* for T cell receptor (TCR) signaling. However, the predictions of such a model derives from logical statements that govern state transitions (e.g., transitions between active and inactive forms of a signaling protein that depend on the activity states of other signaling proteins in a network), and these statements can be viewed as being based on an implicit parameterization.

2. A variety of techniques can be applied to obtain equilibrium constants for protein-protein interactions (Hause et al., 2012; Koytiger et al., 2013) *(56, 57)*, for example, or to determine the overall abundances of proteins in a cell (Kulak et al., 2014; Hein et al., 2015; Shi et al., 2016; Yi et al., 2018) *(58 – 61)*. However, such direct measurements of parameters should be used with caution. Knowing the overall abundance of a protein may not be helpful if the protein is distributed across multiple subcellular compartments. Consider a receptor that is present both at the plasma membrane and within endosomes. In this case, the two populations are distinct: only the population at the plasma membrane is capable of interacting with an extracellular ligand. If ligand-receptor interaction is the primary focus of a model, the abundance of the receptor at the plasma membrane is the parameter of most interest, not the overall abundance. Likewise, knowing the equilibrium constant measured *in vitro* for an isolated protein-protein interaction may not be helpful if the proteins in question interact promiscuously with other binding partners in a cell, because competition will affect the extent to which the binding partners of interest associate (Erickson et al., submitted; Stites et al., 2015) *(46, 62)*. Thus, unless competitive interactions are explicitly considered, the parameter of interest is not the true equilibrium constant but rather the apparent equilibrium constant, i.e., the parameter that implicitly accounts for the effects of competition. For further discussion, see Erickson et al. (submitted) *(46)*. In addition, *in vitro* assays of protein-protein binding that do not involve full-length proteins may miss important allosteric effects, such as autoinhibition.

3. The studies of Shiraishi and Savageau (1992a, 1992b, 1992c) *(63–65)* and Ni and Savageau (1996a, 1996b) *(66, 67)* illustrate the potential problems that can arise when multiple direct *in vitro* measurements of physicochemical parameters are integrated within a mathematical model.

4. A rule-based model is typically written using a formal language akin to a programming language that is designed for the specific purpose of model specification. Thus, rule-based models are similar to programs (Lopez et al., 2013) *(68)*. In contrast, traditional model forms are usually defined in terms of mathematical equations. For recent reviews of rule-based modeling, see Chylek et al. (2013; 2014a; 2014b) *(15, 69–71)* and Stefan et al. (2014) *(72)*.

5. We use the term "biomolecular site" to refer to any functional part of a biomolecule. Thus, in the case of a protein, the term "site" may be taken to refer to a subunit, a domain, a short linear motif (or SLiM), or a particular amino acid residue. An example of a protein site state transition is autophosphorylation of a tyrosine residue within a receptor tyrosine kinase (RTK), such as Y1068 in the epidermal growth factor receptor (EGFR). Another example is binding of the phosphorylated form of this tyrosine residue (pY1068) to a cognate Src homology 2 (SH2) domain, such as that in the adaptor protein GRB2.
6. Smoothing entails multiple stochastic simulation runs and averaging of the results. One should be aware that this brute-force approach for coping with noisy simulation data may not always be appropriate, as when a system has multiple stable steady states with stochastic fluctuations between the basins of attraction. An example of a model with this behavior is that considered in the study of Lipniacki et al. (2008) *(73)*. In this sort of situation, the average of multiple trajectories may not be representative of any trajectory. Smoothing is generally acceptable when the average over multiple stochastic trajectories converges to the deterministic limit. It is a user's responsibility to confirm that smoothing is appropriate for the application at hand.
7. Combinatorial complexity refers to the situation where a small number of biomolecular interactions have the potential to generate a large number of chemical/molecular species because of the combinatorial number of ways that biomolecules may undergo modifications at distinct sites or join together in multicomponent complexes. Combinatorial complexity is a direct consequence of the multisite nature of many types of biomolecules, such as the proteins involved in cell signaling.

8. Although BioNetFit currently interfaces only with the BNGL-compatible simulators available within the BioNetGen framework (including NFsim), users of Kappa may also benefit from BioNetFit because Kappa can be automatically translated to BNGL (and *vice versa*) using TRuML (Suderman and Hlavacek, 2017) *(74)*, which is freely available online (https://github.com/lanl/TRuML) *(75)*.

9. At present, BioNetFit can only consider a single BNGL file per fitting run. There are plans to eliminate this limitation in future releases. In the meantime, we note that multiple models and multiple simulation protocols may be defined using a single BNGL file; we caution that this practice tends to be error prone.

10. Using optimization to find Fourier descriptors is an uncertain and inefficient approach. If one ever encounters a need to solve this type of problem in practice, the methodology of Kuhl and Giardina (1982) *(37)*, for example, should be used instead. Elliptic Fourier analysis (EFA) (Kuhl and Giardina, 1982) *(37)* can be used to find Fourier descriptors for a closed contour directly from a sampling of points from the contour, using whatever number of harmonics $N$ is required for representation to a desired accuracy.

11. EXP files may include information about measurement noise if replicate measurements are available. EXP files of this type are compatible with the chi-square objective function (which may be selected by setting `objfunc=2` in a CONF file). See the BioNetFit user manual *(34)* for details. We do not consider this type of EXP file in this chapter.

12. A dose-response curve reports the value of an experimental readout as a function of some quantity that can be manipulated over a range of values (e.g., the dose of a ligand). Simulation data corresponding to a dose-response curve can be generated using the BNGL command `parameter_scan` (Faeder et al., 2009) *(13)*. This command launches a series of related simulations in which a user-specified parameter is varied systematically in a prescribed manner. For example, a user may request consideration of a specified number of parameter values that lie between specified minimum and maximum values and that are evenly spaced on either a linear or logarithmic scale. Alternatively, a user may request consideration of a specified array of (possibly irregularly spaced) parameter values. Simulation outputs are reported in a SCAN file at a specified time. Often, this report time is chosen to be large such that the outputs characterize steady-state behavior. We will not further consider dose-response data and simulation data from parameter scans. In this chapter, we focus on time-course data. Simulation data corresponding to experimental time-course data can be generated using the BNGL command `simulate`. This command launches a single simulation (of system dynamics). Simulation outputs are reported in a GDAT file at user-specified times.

13. An informal definition of BNGL is given by Faeder et al. (2009) *(13)*. A formal definition is provided as supplementary material in the report of Hogg et al. (2014) *(14)*. BNGL is based on a graphical formalism (Blinov et al., 2006; Lemons et al., 2011) *(28, 76)*.

14. An easy way to create a CONF file for a specific problem is to start with an existing CONF file and modify it for the problem at hand as needed. As noted earlier, CONF files for a collection of example problems are included in the BioNetFit distribution *(29)*.

15. In fitting to pseudo datasets during a bootstrapping job, it is important to randomize the starting point for searches of parameter space. If the starting point is the same for all of the fitting runs, there is a risk of repeatedly converging to a nearby local minimum that is far away from the global minimum. BioNetFit provides multiple options for initializing a search of parameter space. Be careful to select one for which the starting point will vary from fitting run to fitting run. See the BioNetFit user manual for more information *(34)*.

16. A large threshold (e.g., `bootstrap_chi=1e6` for the egg bootstrapping problem, which is a much larger value than the final objective function value for the egg fitting problem) may be specified to minimize or eliminate rejection of fits to resampled data.

# Acknowledgement

Development of BioNetFit has been supported by National Institutes of Health (NIH)/National Institute of General Medical Sciences (NIGMS) grant R01GM111510 (RGP and WSH). EDM acknowledges support from the Laboratory-Directed Research and Development (LDRD) program at Los Alamos National Laboratory (LANL), which is operated for the National Nuclear Security Administration (NNSA) of the US Department of Energy (DOE) under contract DE-AC52-06NA25396. RS is grateful for support from the Los Alamos Center for Nonlinear Studies (CNLS). KEE is supported by the Joint Design of Advanced Computing Solutions for Cancer (JDACS4C) program established by DOE and the National Cancer Institute (NCI) of NIH. JL acknowledges support from the Science, Technology, Engineering, and Mathematics Talent Expansion Program (STEP) of the National Science Foundation (NSF). MRÁ acknowledges support from the DOE/NNSA-sponsored Minority Serving Institutions (MSI) Internship program. The Monsoon cluster at Northern Arizona University (NAU) is supported by the Arizona Board of Regents Technology and Research Initiative Fund (TRIF). The Darwin cluster at Los Alamos National Laboratory is supported by the Computational Systems and Software Environment (CSSE) subprogram of the Advanced Simulation and Computing (ASC) program at LANL, which is funded by DOE/NNSA.

# Figure captions

Fig. 1. Relationships between software components of BioNetGen and BioNetFit and input/output files, which are plain-text files with various filename extensions (as indicated). **(A)** The BioNetGen framework (Blinov et al., 2004; Harris et al., 2016) *(5, 6)* consists of several software components, including BNG2 (written in object-oriented Perl), run_network (written in C/C++), and NFsim (written in C++). RuleBender (written in Java) provides an integrated development environment (IDE) (Xu et al., 2011; Smith et al., 2012) *(8, 9)*, which includes an editor for creating/modifying BNGL files and plotting tools for displaying simulation results stored in GDAT and SCAN files. The run_network and NFsim components implement various simulation algorithms. For example, NFsim (Sneddon et al., 2011) *(7)* implements the network-free stochastic simulation algorithm of Yang et al. (2008) *(26)*, and run_network interfaces with CVODE (Hindmarsh et al., 2005) *(19)*, an ODE solver. After processing a BNGL file, BNG2 launches a simulation by creating a NET or XML file, which is then passed to run_network or NFsim, respectively. The results of a simulation are stored in GDAT and SCAN files. GDAT files store time courses, and SCAN files store dose-response curves. **(B)** BioNetFit (written in Perl) reads BNGL, EXP, CONF, GDAT, and SCAN files and writes BNGL files, as well as various job type-dependent output files. A user-supplied CONF file serves to define a BioNetFit fitting/bootstrapping job and to set the values of parameters of the population-based global metaheuristic optimization algorithm implemented in BioNetFit (Thomas et al., 2016) *(12)*. A set of user-supplied EXP files store the experimental data to be used in fitting. A user-supplied BNGL file defines a model (in the file's model section) and protocols for generating the simulation data to be compared against the experimental data (in the file's actions section). Simulations needed to evaluate goodness of fit are launched by BioNetFit as follows. A temporary BNGL file, derived from the user-supplied BNGL file, is created; this file sets parameters to the current trial values being considered in the optimization algorithm. Next, this file is passed to BioNetGen so as to generate simulation data. The resulting GDAT and/or SCAN files and the user-supplied EXP file(s) are used to determine the value of an objective function, which is defined in the user-supplied CONF file.

Fig. 2. Experimental data for the egg fitting problem. **(A)** A sampling of 180 points from the contour of an egg. These points were sampled from the contour during a constant-velocity traversal, as described in the text. The $(X, Y)$ coordinates of these points are recorded in the EXP file egg.exp. **(B)** Illustration of the format of the EXP file for the egg fitting problem, which is the same as that of a GDAT file. Only selected rows in the file egg.exp, which is available online *(35)*, are shown. The file records two time series: $X(t)$ and $Y(t)$, where each $(X(t), Y(t))$ pair gives the coordinates of the point on the egg's contour that was sampled at time $t$ during the traversal of the contour for $t \in [0,1,...,180]$.

Fig. 3. Listings of the BNGL and CONF files for the egg fitting problem. **(A)** A complete listing of the BNGL file egg.bngl (43 lines). **(B)** A complete listing of the CONF file egg_fit.conf (45 lines).

Fig. 4. The quality of fit for a BioNetFit solution of the egg fitting problem. **(A)** Experimental data (dots) and simulated time courses (curves). The simulated time courses, $X(t)$ (solid curve) and $Y(t)$ (dotted curve), are based on the best-fit parameter values given in **Table 1**, i.e., the Fourier coefficients in **Eqs. 1** and **2** for $N = 2$ found using BioNetFit. Selected $X$ and $Y$ data

points in egg.exp, those at $t = 0, 10, \ldots, 180$, are indicated by dots. **(B)** The solid curve is the representation of the egg's contour obtained using **Eqs. 1** and **2** with best-fit parameter values (**Table 1**). The input contour data (dots) and the EFA representation of the contour (broken curve) are shown for comparison. The dots are identical to the data shown in **Fig. 2.**

Fig. 5. The quality of fit for a BioNetFit solution of the elephant fitting problem. **(A)** Visualization of the input contour data. The contour was sampled, imprecisely, every second during a traversal with period $T = 464$ s. The mean arc length between sampled points is 1.3 (in the same units as *X* and *Y*) with a standard deviation of 0.28. **(B)** EFA representation of the contour using **Eqs. 1** and **2** with $N = 20$. Fourier coefficients were found using the methodology of Kuhl and Giardina (1982) *(37)*. This EFA representation is encoded in the BNGL file elephant_EFA.bngl, which is available online *(36).* **(C)** Representation of the contour based on the best-fit coefficients found using BioNetFit. The fitting job that produced this representation is defined by the files elephant.exp, elephant.bngl, and elephant_fit.conf, which are available online *(36)*. The poor quality of fit illustrates the curse of dimensionality—it can be very difficult, if not impossible, to find the optimum solution in a high-dimensional search space.

# Table captions

Table 1. Summary of the BioNetFit solution to the egg fitting problem. The second column gives Fourier coefficients determined by EFA (Kuhl and Giardina, 1982) *(37)*. The third column gives the best-fit values found via BioNetFit-enabled optimization. The fourth and fifth columns define the 90% confidence intervals found via BioNetFit-enabled bootstrapping. Confidence intervals are based on 1,000 bootstrap samples of the original data.

# Tables

There is 1 table; please see the following page.

Table 1. Summary of the BioNetFit solution to the egg fitting problem. The second column gives Fourier coefficients determined by EFA (Kuhl and Giardina, 1982) *(37)*. The third column gives the best-fit values found via BioNetFit-enabled optimization. The fourth and fifth columns define the 90% confidence intervals found via BioNetFit-enabled bootstrapping. Confidence intervals are based on 1,000 bootstrap samples of the original data.

| Parameter | EFA value | Best-fit value | 90% confidence interval from bootstrapping | |
|---|---|---|---|---|
| | | | Lower limit | Upper limit |
| $a_0$ | 100 | 100 | 99 | 101 |
| $a_1$ | −0.57 | −1.6 | −2.7 | 2.2 |
| $a_2$ | 0.14 | 1.8 | −2.4 | 2.5 |
| $b_1$ | −31 | −30 | −32 | −30 |
| $b_2$ | −1.3 | 1.2 | −3.1 | 2.0 |
| $c_0$ | 140 | 140 | 140 | 140 |
| $c_1$ | −39 | −39 | −40 | −38 |
| $c_2$ | −0.062 | −0.54 | −2.4 | 2.3 |
| $d_1$ | 1.4 | 1.9 | −1.9 | 3.0 |
| $d_2$ | −0.43 | 1.8 | −2.6 | 2.2 |

# Figures

There are 5 figures. The figures are provided as separate files. The filenames are as follows:

- Fig. 1: Fig1_Ch18_Posner.pdf

- Fig. 2: Fig2_Ch18_Posner.pdf

- Fig. 3: Fig3_Ch18_Posner.pdf

- Fig. 4: Fig4_Ch18_Posner.pdf

- Fig. 5: Fig5_Ch18_Posner.pdf

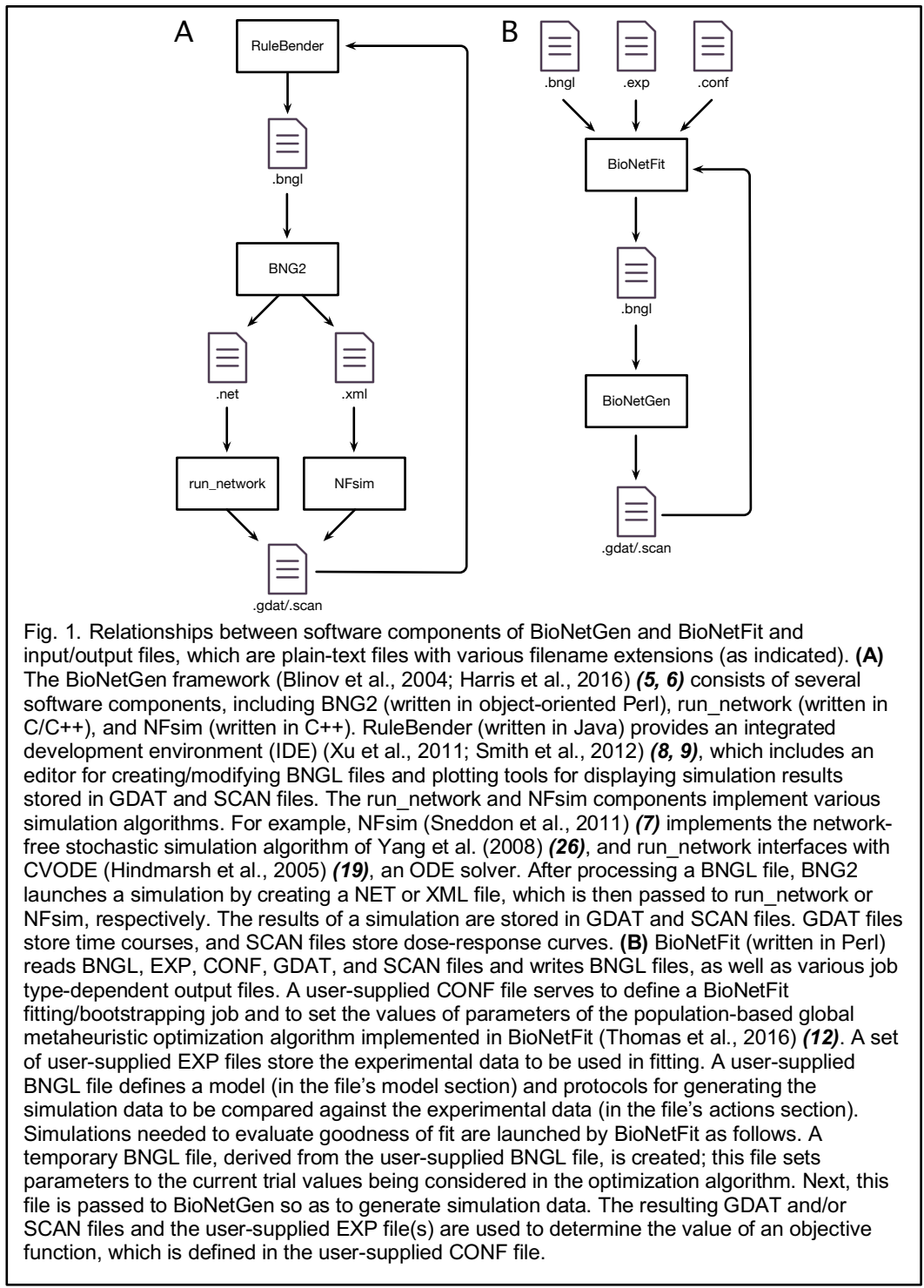

Fig. 1. Relationships between software components of BioNetGen and BioNetFit and input/output files, which are plain-text files with various filename extensions (as indicated). **(A)** The BioNetGen framework (Blinov et al., 2004; Harris et al., 2016) *(5, 6)* consists of several software components, including BNG2 (written in object-oriented Perl), run_network (written in C/C++), and NFsim (written in C++). RuleBender (written in Java) provides an integrated development environment (IDE) (Xu et al., 2011; Smith et al., 2012) *(8, 9)*, which includes an editor for creating/modifying BNGL files and plotting tools for displaying simulation results stored in GDAT and SCAN files. The run_network and NFsim components implement various simulation algorithms. For example, NFsim (Sneddon et al., 2011) *(7)* implements the network-free stochastic simulation algorithm of Yang et al. (2008) *(26)*, and run_network interfaces with CVODE (Hindmarsh et al., 2005) *(19)*, an ODE solver. After processing a BNGL file, BNG2 launches a simulation by creating a NET or XML file, which is then passed to run_network or NFsim, respectively. The results of a simulation are stored in GDAT and SCAN files. GDAT files store time courses, and SCAN files store dose-response curves. **(B)** BioNetFit (written in Perl) reads BNGL, EXP, CONF, GDAT, and SCAN files and writes BNGL files, as well as various job type-dependent output files. A user-supplied CONF file serves to define a BioNetFit fitting/bootstrapping job and to set the values of parameters of the population-based global metaheuristic optimization algorithm implemented in BioNetFit (Thomas et al., 2016) *(12)*. A set of user-supplied EXP files store the experimental data to be used in fitting. A user-supplied BNGL file defines a model (in the file's model section) and protocols for generating the simulation data to be compared against the experimental data (in the file's actions section). Simulations needed to evaluate goodness of fit are launched by BioNetFit as follows. A temporary BNGL file, derived from the user-supplied BNGL file, is created; this file sets parameters to the current trial values being considered in the optimization algorithm. Next, this file is passed to BioNetGen so as to generate simulation data. The resulting GDAT and/or SCAN files and the user-supplied EXP file(s) are used to determine the value of an objective function, which is defined in the user-supplied CONF file.

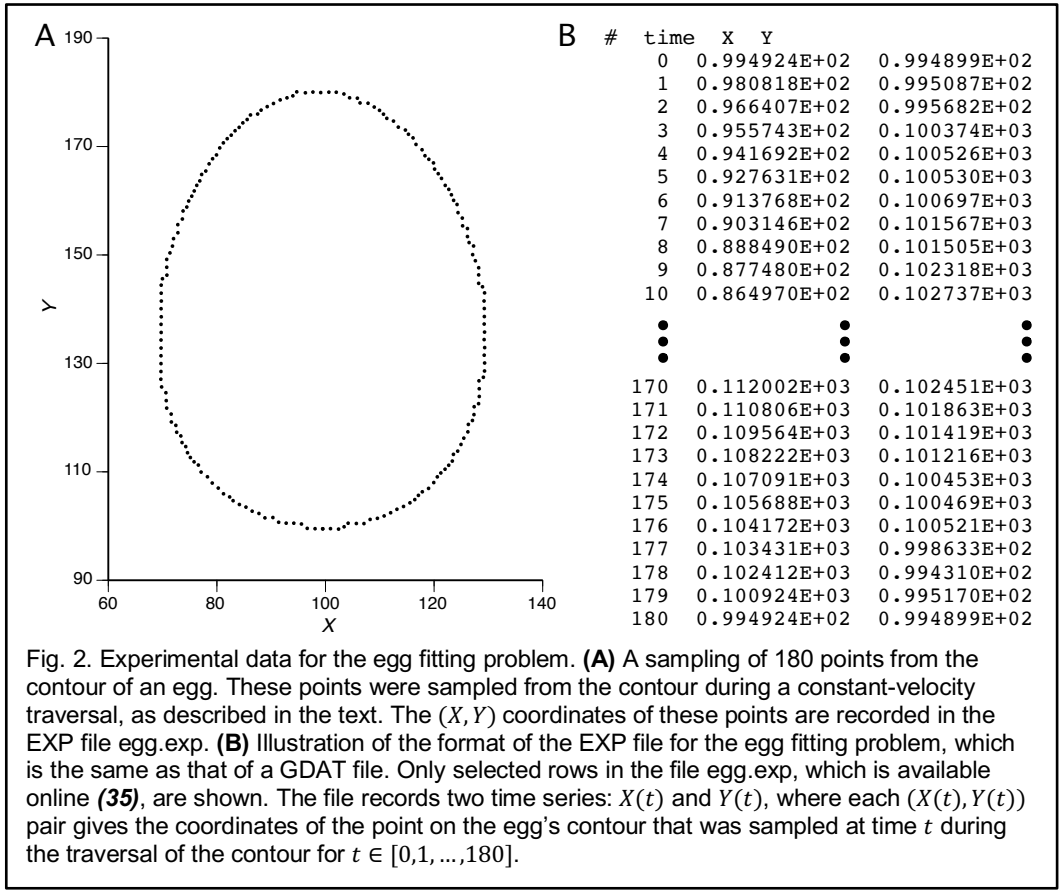

Fig. 2. Experimental data for the egg fitting problem. **(A)** A sampling of 180 points from the contour of an egg. These points were sampled from the contour during a constant-velocity traversal, as described in the text. The $(X, Y)$ coordinates of these points are recorded in the EXP file egg.exp. **(B)** Illustration of the format of the EXP file for the egg fitting problem, which is the same as that of a GDAT file. Only selected rows in the file egg.exp, which is available online *(35)*, are shown. The file records two time series: $X(t)$ and $Y(t)$, where each $(X(t), Y(t))$ pair gives the coordinates of the point on the egg's contour that was sampled at time $t$ during the traversal of the contour for $t \in [0,1,...,180]$.

```
A  1  begin model
   2    begin parameters
   3      a0 a0__FREE
   4      a1 a1__FREE
   5      a2 a2__FREE
   6      b1 b1__FREE
   7      b2 b2__FREE
   8      c0 c0__FREE
   9      c1 c1__FREE
  10      c2 c2__FREE
  11      d1 d1__FREE
  12      d2 d2__FREE
  13      pi=2*asin(1)
  14      period 180
  15      m=2*pi/period
  16    end parameters
  17    begin molecule types
  18      t
  19    end molecule types
  20    begin seed species
  21      t 0
  22    end seed species
  23    begin observables
  24      Species t t
  25    end observables
  26    begin functions
  27      X()=a0\
  28        +a1*cos(m*1*t)+b1*sin(m*1*t)\
  29        +a2*cos(m*2*t)+b2*sin(m*2*t)
  30      Y()=c0\
  31        +c1*cos(m*1*t)+d1*sin(m*1*t)\
  32        +c2*cos(m*2*t)+d2*sin(m*2*t)
  33    end functions
  34    begin reaction rules
  35      0->t 1
  36    end reaction rules
  37  end model
  38  begin actions
  39    generate_network({overwrite=>1})
  40    simulate({suffix=>"egg",method=>"ode",\
  41      t_start=>0,t_end=>180,n_steps=>180,\
  42      print_functions=>1})
  43  end actions
```

```
B  1  # working directory: BioNetFit_v1.01
   2  # input directory for egg fitting job: egg
   3  # command to start fitting job: ./BioNetFit.pl egg/egg_fit.conf
   4
   5  # output directory for fitting/bootstrapping jobs
   6  output_dir=egg_out
   7
   8  # output subdirectory for egg fitting job
   9  job_name=egg_fit
  10
  11  # path to BNG2.pl
  12  bng_command=../RuleBender-2.2.1-osx64/BioNetGen-2.3/BNG2.pl
  13
  14  # path to BNGL file
  15  model=egg/egg.bngl
  16
  17  # path to EXP file
  18  exp_file=egg/egg.exp
  19
  20  # number of processors to use
  21  parallel_count=1
  22
  23  # objective function
  24  objfunc=1
  25
  26  # number of iterations
  27  max_generations=50
  28
  29  # number of trial sets of parameters
  30  #   to be considered in each iteration
  31  permutations=50
  32
  33  # settings for mutation of parameter values
  34  mutate=default 0.2 0.2
  35
  36  # settings for initialization of the search of parameter space
  37  random_var=a0__FREE 95 105
  38  random_var=a1__FREE -50 50
  39  random_var=a2__FREE -50 50
  40  random_var=b1__FREE -50 50
  41  random_var=b2__FREE -50 50
  42  random_var=c0__FREE 135 145
  43  random_var=c1__FREE -50 50
  44  random_var=c2__FREE -50 50
  45  random_var=d1__FREE -50 50
  46  random_var=d2__FREE -50 50
```

Fig. 3. Listings of the BNGL and CONF files for the egg fitting problem. **(A)** A complete listing of the BNGL file egg.bngl (43 lines). **(B)** A complete listing of the CONF file egg_fit.conf (46 lines).

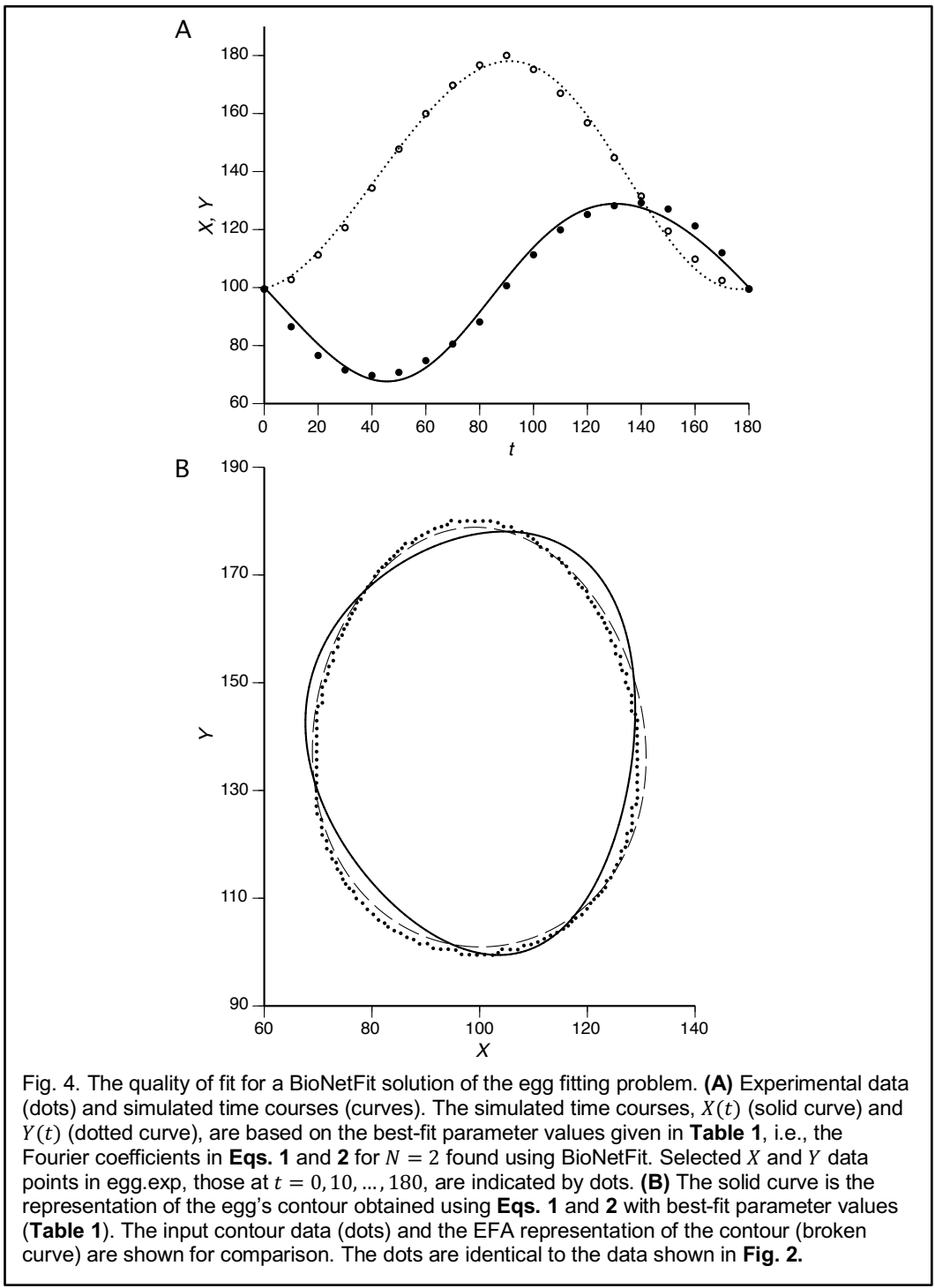

Fig. 4. The quality of fit for a BioNetFit solution of the egg fitting problem. **(A)** Experimental data (dots) and simulated time courses (curves). The simulated time courses, $X(t)$ (solid curve) and $Y(t)$ (dotted curve), are based on the best-fit parameter values given in **Table 1**, i.e., the Fourier coefficients in **Eqs. 1** and **2** for $N = 2$ found using BioNetFit. Selected $X$ and $Y$ data points in egg.exp, those at $t = 0, 10, \ldots, 180$, are indicated by dots. **(B)** The solid curve is the representation of the egg's contour obtained using **Eqs. 1** and **2** with best-fit parameter values (**Table 1**). The input contour data (dots) and the EFA representation of the contour (broken curve) are shown for comparison. The dots are identical to the data shown in **Fig. 2.**

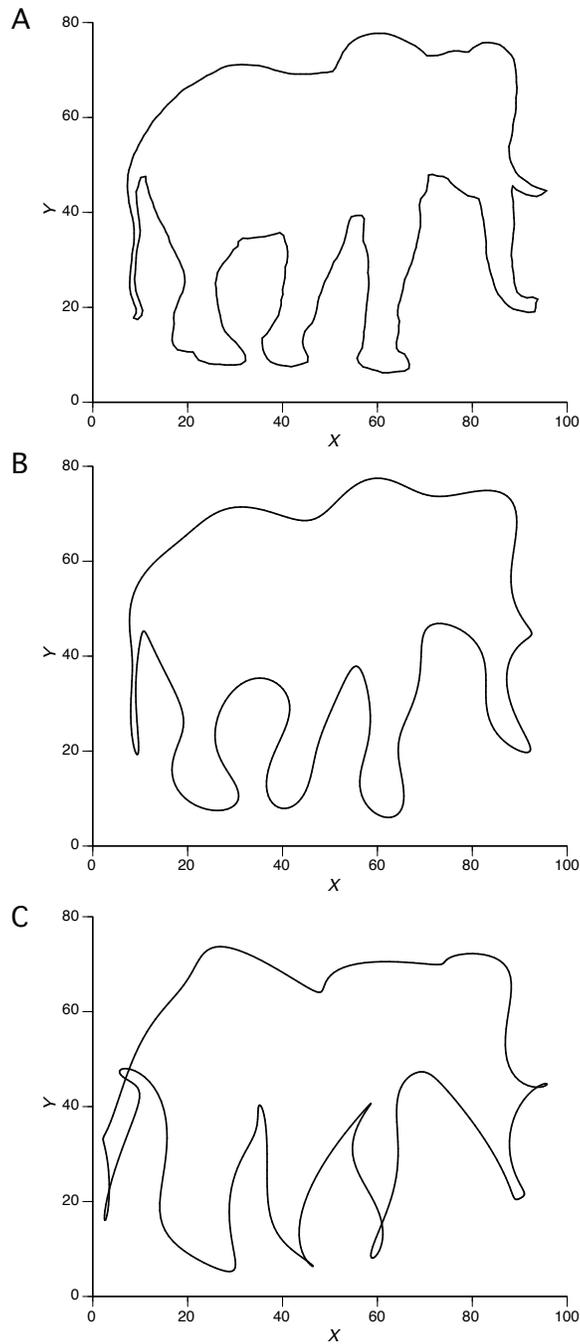

Fig. 5. The quality of fit for a BioNetFit solution of the elephant fitting problem. **(A)** Visualization of the input contour data. The contour was sampled, imprecisely, every second during a traversal with period $T = 464$ s. The mean arc length between sampled points is 1.3 (in the same units as *X* and *Y*) with a standard deviation of 0.28. **(B)** EFA representation of the contour using **Eqs. 1** and **2** with $N = 20$. Fourier coefficients were found using the methodology of Kuhl and Giardina (1982) *(37)*. This EFA representation is encoded in the BNGL file elephant_EFA.bngl, which is available online *(36)*. **(C)** Representation of the contour based on the best-fit coefficients found using BioNetFit. The fitting job that produced this representation is defined by the files elephant.exp, elephant.bngl, and elephant_fit.conf, which are available online *(36)*. The poor quality of fit illustrates the curse of dimensionality—it can be very difficult, if not impossible, to find the optimum solution in a high-dimensional search space.